\documentclass[10pt, conference]{IEEEtran}

\usepackage{amsmath}
\usepackage{amssymb}
\usepackage{verbatim}
\usepackage{epsfig}

\newtheorem{lemma}{Lemma}
\newtheorem{conjecture}{Conjecture}

\newcommand{\bh}{{\bf h}}

\newcommand{\bw}{{\bf w}}

\newcommand{\bH}{{\bf H}}
\newcommand{\bX}{{\bf X}}
\newcommand{\bQ}{{\bf Q}}
\newcommand{\bx}{{\bf x}}

\newcommand{\bq}{{\bf q}}
\newcommand{\bv}{{\bf v}}

\newcommand{\bheff}{{\bf h}^{\textrm{eff}}}
\newcommand{\yeff}{y^{\textrm{eff}}}

\begin{document}

\title{A Feedback Reduction Technique for \\ MIMO Broadcast Channels}

\author{\authorblockN{Nihar Jindal}
\authorblockA{Department of Electrical and Computer Engineering\\
University of Minnesota\\
Minneapolis, MN 55455, USA \\
Email: nihar@umn.edu}}

\maketitle

\begin{abstract}
A multiple antenna broadcast channel with perfect channel
state information at the receivers is considered.  If each receiver
quantizes its channel knowledge to a finite number of bits 
which are fed back to the transmitter, the large
capacity benefits of the downlink channel can be realized.
However, the required number of feedback bits per mobile 
must be scaled with both the number of transmit antennas and the
system SNR, and thus can
be quite large in even moderately sized systems.  It is shown
that a small number of antennas can be used at each
receiver to improve the quality of the channel estimate 
provided to the transmitter.  As a result, the required
feedback rate per mobile can be significantly decreased.

\end{abstract}

\section{Introduction}

In multiple antenna broadcast (downlink) channels, capacity can
be tremendously increased by adding antennas
at only the access point (transmitter) 
\cite{Caire_Shamai}\cite{Jindal_Goldsmith_DPC}.
However, the transmitter must
have accurate channel state information (CSI) in order to
realize these multiplexing gains.  In frequency-division duplexed
systems, training can be used to obtain channel knowledge at
each of the mobile devices (receivers), but obtaining CSI at the access
point generally requires feedback from each mobile.

In the practically motivated \textit{finite rate feedback} model, 
each mobile feeds back a finite number of bits regarding its 
channel instantiation at the beginning of each block or frame.
The feedback bits are determined by quantizing the channel vector
to one of $2^{B}$ quantization vectors.
A downlink channel with such a feedback mechanism was
analyzed in \cite{Jindal_Limited_FB}\cite{Ding_Love_Zoltowski}.
While only a few feedback bits 
suffice to obtain 
near-perfect CSIT performance in point-to-point MISO 
(multiple-input, single-output)
channels \cite{Santipach_Honig}\cite{Love_Heath_Santipach_Honig},
considerably more feedback is required in downlink channels.
In fact, the feedback load per mobile must be scaled with the
number of transmit antennas as well as the system SNR in downlink channels
in order to achieve rates close to those achievable with perfect CSIT.  
In \cite{Jindal_Limited_FB}, it is shown that the following 
scaling of feedback bits per mobile
\begin{eqnarray} \label{eq-scaling}
B = \frac{M-1}{3} P_{dB}
\end{eqnarray}
suffices to maintain a maximum gap of 3 dB between 
perfect CSIT and limited feedback performance.
This feedback load, however, can be prohibitively large for even reasonable
size systems.  In a 10 antenna system operating at 10 dB, for example, 
this equates to 30 feedback bits per mobile. 

In this paper, we propose a method that significantly reduces the required
feedback load by utilizing a small number of receive antennas 
(denoted by $N$) at each mobile.
The multiple receive antennas are not used to
increase the number of data streams received at each mobile, as they
are in point-to-point MIMO systems, but instead are used to
improve the quality of the channel estimate provided to the transmitter.
Each mobile linearly combines the received
signals on its $N$ antennas to produce a scalar output,
thereby creating an effective single antenna channel at each mobile.
Transmission is then performed as in a multiple transmit
antenna, single receive antenna downlink channel.
However, the coefficients of the linear combiner at each mobile are not
arbitrary, but instead are chosen to produce the effective 
single antenna channel that can be quantized with
minimal error, thereby decreasing the quantization error for each 
mobile.
Increasing the number of receive antennas $N$ clearly
increases the space of possible effective channels, and thus
leads to reduced quantization error.  In a 10 antenna
system operating at 10 dB, for example, 
this method reduces the feedback
from 30 bits per mobile in the $N=1$ scenario to
25 bits and 21 bits, for $N=2$ and $N=3$, respectively.

Notation: We use lower-case boldface to denote vectors,
upper-case boldface for matrices, and the symbol $(\cdot)^H$
for the conjugate transpose.
The norm of vector ${\mathbf x}$ is denoted $|| {\mathbf x} ||$.

\section{System Model}

We consider a $K$ receiver multiple antenna broadcast
channel in which the transmitter (access point) has
$M$ antennas, and each of the receivers has $N$ antennas.
The received signal at the $i$-th antenna is described as:
\begin{equation}
y_i = {\mathbf h}_i^H {\mathbf x} + n_i, ~~~
   i=1,\ldots,NK
\end{equation}
where ${\bf h}_1, {\bf h}_2, \ldots, {\bf
        h}_{KN}$ are the channel vectors (with ${\bf h}_i \in
        {\mathbb{C}}^{M \times 1}$) describing the $KN$
receive antennas, the vector 
${\bf x} \in {\mathbb{C}}^{M \times 1}$  
is the transmitted signal, and ${\bf n}_1, \ldots, {\mathbf n}_{NK}$
are independent complex Gaussian noise terms
with unit variance.
Note that receiver 1 has access to signals $y_1, \ldots, y_N$,
receiver 2 has access to $y_{N+1}, \ldots, y_{2N}$, and
the $i$-th receiver has access to $y_{(i-1)N+1},\ldots, y_{Ni}$.
There is a transmit power constraint of $P$, i.e.
we must satisfy $E[||{\mathbf x}||^2] \leq P$.
We use $\bH_i$ to denote the concatenation of the
$i$-th receiver's channels, i.e. 
$\bH_i = [\bh_{(i-1)N+1} \cdots \bh_{Ni}]$.
For simplicity of exposition we assume that
the number of mobiles is equal to the number of transmit
antennas, i.e., $K=M$.  The results can easily
be extended to the case where $K < M$, and the proposed
technique can be combined with user selection when $K > M$.
Furthermore, the number of receive antennas is assumed
to be no larger than the number of transmit antennas.


We consider a block fading channel, with independent
Rayleigh fading from block to block. Each of the receivers
is assumed to have perfect and instantaneous knowledge of
its own channel $\bH_i$.  Notice it is not
required for mobiles to know the channel of other mobiles.
In this work we study only the \textit{ergodic capacity}, or the
average rates achieved over an infinite number of blocks (or
channel realizations).

\subsection{Finite Rate Feedback Model} \label{sec-finite}
Here we briefly describe the feedback model for a
single receive antenna ($N=1$).  At the beginning of each block,
each receiver quantizes its channel 
(with ${\bf h}_i$ assumed to be known
perfectly at the $i$-th receiver) to $B$ bits and feeds back
the bits perfectly and instantaneously to the access point.
Vector quantization is performed using a codebook ${\mathcal C}$ 
that consists of $2^{B}$ $M$-dimensional unit norm vectors
${\mathcal C} \triangleq \{ \mathbf{w}_1, \ldots, \mathbf{w}_{2^{B}} \}$,
where $B$ is the number of feedback bits.
Each receiver quantizes its channel vector
to the beamforming vector that forms the minimum angle to it,
or equivalently that maximizes the inner product
\cite{Love_Heath} \cite{Mukkavilli_MIMO}.
Thus, user $i$ quantizes its channel to $\hat{\bh}_i$, chosen according to:
\begin{eqnarray} \label{eq-quant}
\hat{\bh}_i &=& \textrm {arg} \max_{\bw=\bw_1,\ldots,\bw_{2^{B}}} | {\mathbf h}_i^H \bw | \\
&=& \textrm {arg} \min_{\bw = \bw_1,\ldots,\bw_{2^{B}}}
\sin^2 \left( \angle ({\bf h}_i, \bw) \right).
\end{eqnarray}
and feeds the index of the quantization back to the transmitter.
It is important to notice that only the direction of the channel vector 
is quantized, and no magnitude information 
is conveyed to the transmitter.  The quantization
error can be thought of as either the angle between the channel
and its quantization $\angle ({\bf h}_i, \hat{\bh}_i)$
or the quantity $\sin^2 (\angle ({\bf h}_i, \hat{\bh}_i))$.

In this work we use \textit{random vector quantization} (RVQ),
in which each of the $2^{B}$ quantization
vectors is independently chosen from the isotropic distribution
on the $M$-dimensional unit sphere \cite{Santipach_Honig}.  
To simplify analysis, each
receiver is assumed to use a different and independently generated
codebook.  We analyze performance
averaged over all such choices of random codebooks.
Random codebooks are used
because the optimal vector quantizer for this problem is
not known in general and known bounds are rather loose,
whereas RVQ is amenable to analysis and also provides
performance that is measurably close to the optimal \cite{Santipach_Honig}. 

\subsection{Zero-Forcing Beamforming} \label{sec-zfbf}
After receiving the quantization indices from each of the
mobiles, the AP can use zero-forcing beamforming (ZFBF) to transmit
data to the $M$ users.  Let us again consider the $N=1$ scenario,
where the channels are the vectors $\bh_1, \ldots, \bh_M$.
Since the transmitter does not have perfect CSI,
ZFBF must be performed based on the quantizations instead of
the actual channels.  When ZFBF is used, the transmitted signal
is defined as $\bx = \sum_{i=1}^M x_i \bv_i $, where each $x_i$ is a scalar
(chosen complex Gaussian with power $P/M$) intended for the $i$-th receiver, 
and $\bv_i \in {\mathcal C}^M$ is the beamforming
vector for the $i$-th receiver.  The beamforming vectors
$\bv_1, \ldots, \bv_M$ are chosen as the normalized rows
of the matrix $[ \hat{\bh}_i \cdots \hat{\bh}_M]^{-1}$, i.e.,
they satisfy $||\bv_i|| = 1$ for all $i$ 
and  $\hat{\bh}_i^H \bv_j = 0$ for all $j \ne i$.
If all multi-user interference is treated as additional noise,
the resulting SINR at the $i$-th receiver is given by:
\begin{eqnarray} \label{eq-downlink_sinr}
SINR_i = \frac { \frac{P}{M} | \bh_i^H \bv_i |^2 }
{1 + \sum_{j \ne i}  \frac{P}{M} | \bh_i^H \bv_j |^2 }.
\end{eqnarray}
Note that the interference terms in the denominator are
strictly positive because $\bh_i \ne \hat{\bh}_i$, i.e., due
to the quantization error.

\section{Effective Channel Quantization} \label{sec-quantization}
In this section we describe the proposed method to
reduce the quantization error in the transmitter's estimate
of the mobile channels.  We begin by first 
describing a simple, antenna-selection method
for reducing feedback, which motivates the better performing
effective channel method.

A simple method to utilize $N$ receive antennas is 
to separately quantize each of the $N$ channel vectors
and then feed back the index of only the best of the 
$N$ quantizations.
If, for example, antenna $1$ had the minimum quantization error,
the mobile would only send the quantization index describing
antenna $1$ and would only utilize the first antenna when
receiving.  It is straightforward to show
that choosing the best of the $N$ channel quantizations, 
each from  a codebook of size $2^{B}$, is
equivalent to quantizing a single channel using a codebook
of size $N \cdot 2^{B}$.
Thus, if $B$ feedback bits are sent by each mobile,
a system with $N$ antennas per mobile will perform identical
to a single receive antenna system with $B + \log_2 N$
feedback bits per mobile.  Thus, utilizing $N$ receive antennas
in this simple manner decreases the feedback load by
$\log_2 N$ bits per mobile.


A more significant decrease in feedback load can be
achieved by considering all possible \textit{linear combinations }
of the $N$ received signals, instead of limiting
the system to selection of one of the $N$ signals.  Consider
the effective received signal at the first receiver after
linearly combining the $N$ received signals by complex weights
$\boldsymbol{\gamma}_1=(\gamma_{1,1}, \ldots, \gamma_{1,N})$ 
satisfying $|\boldsymbol{\gamma}_1|=1$:
\begin{eqnarray*}
\yeff_1 = \sum_{k=1}^N \gamma_{1,k} y_k &=&
\sum_{k=1}^N  \gamma_{1,k}(\bh_k^H \bx + n_k) \\
&=& \left( \sum_{k=1}^N  \gamma_{1,k}\bh_k^H \right) \bx + n \\
&=& (\bheff_1)^H  {\bf x} + n,
\end{eqnarray*}
where $\bheff_1 = \sum_{k=1}^N  \gamma_{1,k}\bh_k = 
\bH_1 {\boldsymbol{\gamma}}_1$ and
$n = \sum_{k=1}^N  \gamma_{1,k}n_k$ is unit variance complex Gaussian noise
because $|\boldsymbol{\gamma}_1|=1$.
Since any set of weights satisfying the unit norm can be chosen,
$\bheff_1$ can be in \textit{any} direction in the
subspace spanned by $\bh_1, \ldots, \bh_N$.
Thus, the quantization error is minimized by choosing $\bheff_1$ 
to be in the direction that can be quantized best, or equivalently the
direction which is closest to one of the quantization vectors.



Let us now more formally describe the quantization process
performed at the first mobile.
As described in Section \ref{sec-finite}, the quantization codebook 
consists of $2^{B}$ isotropically
chosen unit norm vectors $ \mathbf{w}_1, \ldots, \mathbf{w}_{2^{B}}$.
In the single receive antenna ($N=1, ~ \bH_i = \bh_i$) 
scenario, the best quantization
corresponds to the vector maximizing $|{\bf h}_1^H {\bf w}_j|$.
Since $|{\bf h}_1^H {\bf w}_j| = ||{\bf h}_1|| \cdot
|\cos (\angle(\bh_1, {\bf w}_j)|$, this is equivalent to choosing the
quantization vector that has the smallest angle between itself
and the channel vector $\bh_1$.  When $N>1$, we compute the angle
between each quantization vector and the \textit{subspace} spanned
by the $N$ channel vectors, and pick the quantization vector that forms
 the smallest such angle.  Alternatively, each quantization vector is
projected onto the span of the $N$ channel vectors, and the
angle between the quantization vector and its projection is
computed.  If ${\bf q}_1, \ldots, {\bf q}_N$ 
forms an orthonormal basis for span$(\bh_1, \ldots, \bh_N)$
(easily computable using Gram-Schmidt), then
the quantization is performed according to:
\begin{eqnarray}
\hat{\bh}_1 &=& \textrm {arg} \min_{\bw = \bw_1,\ldots,\bw_{2^{B}}}
| \angle( \bw, \textrm{span}({\bf h}_1, \ldots, {\bf h}_N)) | \\
&=& \textrm {arg}
\max_{\bw = \bw_1,\ldots,\bw_{2^{B}}} \sum_{k=1}^N |\bw^H {\bf q}_k|^2.
\end{eqnarray}
Let us denote the normalized projection of $\hat{\bh}_1$
onto span$({\bf h}_1, \ldots, {\bf h}_N)$ by the vector
${\bf s}_1^{\textrm{proj}}$.  Notice that the
direction specified by  ${\bf s}_1^{\textrm{proj}}$
has the minimum quantization error amongst all
directions in span$({\bf h}_1, \ldots, {\bf h}_N)$.

Next we describe the method used to
choose the $N$-dimensional weight vector ${\boldsymbol \gamma}_1$. 
We wish to choose a unit norm vector ${\boldsymbol \gamma}_1$ such that
$\bheff=\sum_{j=1}^N \gamma_{1,j} {\bf h}_j = \bH_1 {\boldsymbol \gamma}_1$
 is in the direction of the projected quantization vector 
${\bf s}_1^{\textrm{proj}}$.
First we find the vector $\bf v \in \mathcal{C}^N$ such that
${\bf H}_1 {\bf v} = {\bf s}_1^{\textrm{proj}}$, and then scale to
get  ${\boldsymbol \gamma}_1$.
Since ${\bf s}_1^{\textrm{proj}}$ is in span$(\bH_1)$, 
${\bf v}$ can be found by the pseudo-inverse of $\bH_1$:
\begin{eqnarray}
{\bf v} = \left( {\bf H}_1^H {\bf H}_1 \right)^{-1} 
{\bf H}_1^H {\bf s}_1^{\textrm{proj}},
\end{eqnarray}
and the coefficient vector  ${\boldsymbol \gamma}_1$ is the
normalized version of ${\bf v}$: 
$\boldsymbol{\gamma} = \frac { {\bf v}} { ||{\bf v}|| }$.
It is easy to check that $||\bheff_1 || =  1 / || {\bf v} ||$.

\begin{figure}
    \centering
    \epsfig{file=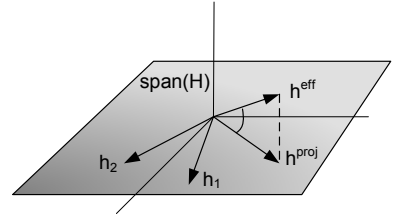, width=2in}
\caption{Quantization procedure for a two antenna mobile}
   \label{fig-projection}
\end{figure}

The quantization procedure is illustrated for a $N=2$ channel in
Fig. \ref{fig-projection}.  In the figure the span of the two channel vectors is
shown, along with the projection of the best quantization vector
onto this subspace along with the subsequent angular error.

We now summarize the procedure for computing the quantization
vector and the weighting vector of the $i$-th mobile:
\begin{enumerate}

\item Compute the channel quantization:
\begin{eqnarray} \nonumber
\hat{\bh}_i &=& \textrm {arg} \min_{\bw = \bw_1,\ldots,\bw_{2^{B}}}
| \angle( \bw, \textrm{span}(\bH_i)) | \\
&=& \textrm {arg} \max_{\bw = \bw_1,\ldots,\bw_{2^{B}}}
\sum_{k=1}^N |\bw^H {\bf q}_k|^2.
\end{eqnarray}
where ${\bf q}_1, \ldots, {\bf q}_N$ is an orthonormal
basis for the span of the columns of $\bH_i$.

\item
Project the quantization vector onto the span of the channel
vectors:
\begin{eqnarray*}
{\bf s}_i^{\textrm{proj}} = \frac{ \sum_{k=1}^N
{\bf q}_k ( \hat{\bh}_i^H {\bf q}_k) }
{|| \sum_{k=1}^N {\bf q}_k (\hat{\bh}_i^H {\bf q}_k)|| }.
\end{eqnarray*}

\item 
Compute the weighting vector ${\boldsymbol \gamma}_i$:
\begin{eqnarray} \label{eq-coeff}
{\boldsymbol \gamma}_i = \frac { \left( {\bf H}_i^H {\bf H}_i \right)^{-1} {\bf H}_i^H {\bf s}_1^{\textrm{proj}} } 
{ || \left( {\bf H}_i^H {\bf H}_i \right)^{-1} {\bf H}_i^H {\bf s}_1^{\textrm{proj}} || }.
\end{eqnarray}
\end{enumerate}

Each mobile performs these steps, feeds back the index
of its quantized channel, 
and then linearly combines its $N$ received signals
using weighting vector ${\boldsymbol \gamma}_i$
to get $y^{\textrm{eff}}_i = (\bheff_i)^H {\bf x} + n$
with $\bheff_i = \bH_i {\boldsymbol \gamma}_i$.

The proposed method finds the effective channel with the
minimum quantization error without any regard to the resulting
channel magnitude (i.e., $||\bheff_i||$).  This is 
reasonable because quantization error is typically the
dominating factor in  limited feedback downlink systems, as we
later see in the sum rate analysis.
However, it may be useful to later study alternatives that balance
minimization of quantization error with maximization of 
channel magnitude.

\section{Sum Rate Analysis}
The effective channel quantization procedure converts the
multiple transmit, multiple receive antenna downlink channel
into a multiple transmit, single receive antenna downlink channel
with channel vectors $\bheff_1, \ldots, \bheff_M$ and
channel quantizations $\hat{\bh}_i \cdots \hat{\bh}_M$. 
In fact, the transmitter need not even be aware of the number of 
receive antennas, since the multiple receive antennas are 
used only during quantization.

After receiving the quantization indices from each of the
mobiles, the transmitter performs zero-forcing beamforming
(as described in Section \ref{sec-zfbf}) based on the channel
quantizations. The resulting SINR at the $i$-th receiver is given by:
\begin{eqnarray}  \label{eq-sinr_eff}
SINR_i = \frac { \frac{P}{M} | (\bheff_i)^H \bv_i |^2 }
{1 + \sum_{j \ne i}  \frac{P}{M} | (\bheff_i)^H \bv_j |^2 }.
\end{eqnarray}
We are interested in the long-term average sum rate
achieved in this channel, and thus the expectation of
$\sum_{i=1}^M \log(1 + SINR_i)$.  Since the
beamforming vectors are chosen according to the
ZFBF criterion based on the quantized channels, they
satisfy $||\bv_i|| = 1$ for all $i$ 
and  $\hat{\bh}_i \bv_j = 0$ for all $j \ne i$.
Quantization error, however, leads to mismatch
between the effective channels and their quantizations, and thus
strictly positive interference terms (of the form $| (\bheff_i)^H \bv_j |^2$)
in the denominator of the SINR expression.

\subsection{Preliminary Calculations} 
In order to analyze the expected rate of such a system, 
the distribution of the quantization error (between
$\bheff_i$ and $\hat{\bh}_i$) and of the effective channel
must be characterized.  Note that we
consider these distributions over the randomly generated
channels as well as the random vector quantization.

\begin{lemma} \label{lemma-quant_error}
The quantity $\cos^2 ( \angle( \hat{\bh_i}, \bheff_i ))$, 
which is one minus the quantization error 
$\sin^2 ( \angle( \hat{\bh_i}, \bheff_i ))$, 
is the maximum of
$2^B$ independent beta $(N,M-N)$ random variables.
\end{lemma}
\begin{proof}
Let ${\bf q}_1, \ldots, {\bf q}_N$ represent an orthonormal
basis for span$(\bH_i)$.  Since the quantization vectors are
unit norm, $\cos^2 \left(\angle( {\bf w}_j, \textrm{span}(\bH_i) \right)
= \sum_{k=1}^N |{\bf w}_j^H {\bf q}_k|^2$
for any quantization vector.
Since the basis vectors and quantization vectors are isotropically chosen,
this quantity is the squared norm
of the projection of a random vector in ${\mathcal C}^M$ 
onto a random $N$-dimensional subspace, which is described by the
beta distribution with parameters $N$ and $M-N$ \cite{Beta}.  Furthermore,
these random variables are independent for different quantization vectors
(i.e., different $j$) due to the independence of the 
quantization vectors and channels.
\end{proof}
Using the basic properties of the beta distribution, this
implies that the quantization error
which is one minus the quantization error 
$\sin^2 ( \angle( \hat{\bh_i}, \bheff_i ))$, 
is the minimum of $2^B$ independent beta $(M-N,N)$ random variables.

The following lemma and conjecture characterize the distribution
of the effective channel vectors.
\begin{lemma} \label{lemma-effective_isotropic}
The normalized effective channels $\frac{\bheff_1}{||\bheff_1||}, \ldots, 
\frac{\bheff_M}{||\bheff_M||}$ are iid isotropic vectors in 
${\mathcal C}^M$.
\end{lemma}
\begin{proof}
From the earlier description of effective channel quantization,
note that $\frac{\bheff_i}{||\bheff_i||} = {\bf s}_i^{\textrm{proj}}$,
which is the projection of the best quantization vector
onto span$(\bH_i)$.  Since each quantization vector is chosen
isotropically, its projection 
is isotropically distributed within the subspace.  Furthermore,
the best quantization vector is chosen  based solely on the angle 
between the quantization vector and its projection.  Thus
${\bf s}_i^{\textrm{proj}}$ is isotropically distributed
in span$(\bH_i)$.  Since this subspace is also isotropically
distributed, the vector ${\bf s}_i^{\textrm{proj}}$ is
isotropically distributed in ${\mathcal C}^M$.
To show independence, note that the quantization vectors
and the channel realizations, from which the effective channel
is generated, are independent from mobile to mobile.
\end{proof}

\begin{conjecture} \label{conj}
The squared norm of the effective channel $\bheff_i$
is chi-squared with $2(M-N+1)$ degrees of freedom.
\end{conjecture}

While this conjecture can be proven for the case when
$N=M$ using the fact that the diagonal entries of
$(\bH_i \bH_i^H)^{-1}$ are each inverted chi-square with
two degrees of freedom when $\bH_i$ is square 
\cite[Theorem 3.2.12]{Muirhead},
this proof does not yet extend to the scenario where
$1 < N < M$.  However, numerical results very strongly indicate that
the conjecture is true for all values of $N$ and $M$.  The claim is
trivially true when $N=1$ because $\bheff_i = \bh_i$ when
mobiles have a single antenna).  Furthermore, it
is known that $1/( \bv^H \bH_i^H \bH_i)^{-1} \bv)$
is chi-square distributed with  $2(M-N+1)$ degrees of freedom
for any unit norm $\bv$ \cite{Muirhead}.

Note that if $N = M$, there is zero quantization error but the
resulting effective channels have only two degrees of freedom.
This scenario is not relevant, however, 
because higher rates can be achieved by
simply transmitting to a single user using point-to-point
MIMO techniques, since such a system has the same
number of spatial degrees of freedom as the downlink channel.
If $N < M$, the quantization error is strictly positive
with probability one by the properties of the beta distribution.

\subsection{Sum Rate Performance Relative to Perfect CSIT}

In order to study the effect of finite rate feedback,
we compare the sum rate achieved using finite rate feedback and
effective channel quantization (for $N \geq 1$), denoted
$R_{FB}(P)$, to the sum rate achieved with perfect CSIT in an $M$ transmit,
\textit{single} receive antenna downlink channel, denoted $R_{ZF}(P)$.
We use the single receive antenna downlink with perfect CSIT
as the benchmark instead of the $N$ receive antenna perfect CSIT 
downlink channel because the proposed method effectively
utilizes a single receive antenna per mobile for reception, and thus
cannot outperform a single receive antenna downlink channel with perfect
CSIT, even in the limit of an infinite number of feedback bits.
Furthermore, this analysis allows us to compare the required feedback
load with $N > 1$ and the proposed method to the
required feedback load for downlink channels with single receive
antennas, studied in \cite{Jindal_Limited_FB}.

Let us first analyze the rates achieved in a single receive
antenna downlink channel using ZFBF under the assumption of
perfect CSIT. If the transmitter
has perfect CSIT, the beamforming vectors (denoted ${\bf v}_{ZF,i}$) 
can be chosen perfectly orthogonal
to all other channels, thereby eliminating all multi-user interference.
Thus, the SNR of each user is as given in (\ref{eq-downlink_sinr})
with zero interference terms in the denominator.
The resulting average rate is given by:
\begin{eqnarray*}
E_{{\bf H}} [R_{ZF}(P)] =  E_{{\bf H}} \left[ \log \left(1 + \frac{P}{M} |{\bf h}_i^H
{\bf v}_{ZF,i}|^2 \right) \right].
\end{eqnarray*}
Since the beamforming vector ${\bf v}_{ZF,i}$ is chosen
orthogonal to the $(M-1)$ other channel vectors $\{ \bh_j \}_{j \ne i}$,
each of which is an iid isotropic vector, the beamforming vector
is also an isotropic vector, \textit{independent}
of the channel vector $\bh_i$.  Because the effective
channel vectors are isotropically distributed (Lemma 
\ref{lemma-effective_isotropic}), the same is true of the
beamforming vectors and the effective channel vectors when the
proposed method is used.

If the number of feedback bits is fixed, the rates achieved with
finite rate feedback are bounded even as the SNR is taken
to infinity.  Thus, the number of feedback bits must be appropriately
scaled in order to avoid this limitation.  Furthermore, it is
useful to consider the scaling of bits required to maintain a desired
rate (or power) gap between perfect CSIT and limited feedback.  Thus,
we study the rate gap at asymptotically high SNR, denoted as $\Delta R$.
Some simple algebra yields the following upper bound to $\Delta R$:
\begin{eqnarray*}
\Delta R &\triangleq& \lim_{P \rightarrow \infty} 
E_{{\bf H},W} [R_{ZF}(P) - R_{LF}(P)] \nonumber \\
&\leq& \lim_{P \rightarrow \infty} 
E_{{\bf H}} \left[ \log \left(1+\frac{P}{M} |{\bf h}_i^H
{\bf v}_{ZF}|^2 \right) \right] -  \\ &&
E_{{\bf H},W} \left[ \log \left(1+\frac{P}{M} | (\bheff_i)^H {\mathbf v}_i |^2  \right) \right] +  \\ 
&& E_{{\bf H},W} \left[ \log \left(
1 + \sum_{j \ne i}  \frac{P}{M} | (\bheff_i)^H {\mathbf v}_j |^2 \right) 
\right] 
\end{eqnarray*}
The difference of the first two terms, 
which is the rate loss due to the reduced effective channel
norm (Conjecture \ref{conj}), can be computed in closed form
using the expectation of the log of chi-square random variables, giving
a loss of $\Delta_a \triangleq \log_2 e \sum_{l=M-N+1}^{M-1} \frac{1}{l}$.
The final term, which is the rate loss due to the quantization error,
 can be upper bounded using Jensen's inequality and some of the 
techniques from \cite{Jindal_Limited_FB} to give:
\begin{eqnarray*}
\Delta R &\leq&  \Delta_a + 
\log E_{{\bf H},W} \left[
1 + \sum_{j \ne i}  \frac{P}{M} | (\bheff_i)^H {\mathbf v}_j |^2 \right] \\
&=&  \Delta_a + 
\log \left[ 1\! +\! P\! \left( \frac{M\!-N\!+1}{M} \right) \!
E[\sin^2 ( \angle( \hat{\bh_i}, \bheff_i ))]
 \right]
\end{eqnarray*}

We now utilize Lemma \ref{lemma-quant_error} to estimate the quantization
error.  If we let $X$ be a beta$(M-N, N)$ random variable, the 
CCDF of $X$ can be accurately approximated for $x \approx 1$ as
$\textrm{Pr}(X \geq x) \approx 
{M \! - \! 1 \choose N \! - \! 1} (1-x)^{(M-N)}$.
Since the quantization error is one minus the maximum of 
$2^B$ such random variables,
we use extreme value theory and find $x$ such that 
$\textrm{Pr}(X \geq x) = 2^{-B}$ to get the following approximation for the 
quantization error:
\begin{eqnarray*}
E[\sin^2 ( \angle( \hat{\bh_i}, \bheff_i ))] \approx 2^{-\frac{B}{M-N}}
{M \! - \! 1 \choose N \! - \! 1}^{-\frac{1}{M-N}}
\end{eqnarray*}
Thus we have
\begin{eqnarray*}
\Delta R &\approx&  \log_2 e \sum_{l=M-N+1}^{M-1} \frac{1}{l} + \\
&& \log \left( 1 \! + \! P \cdot \left( \frac{M \! - \! N \! + \! 1}{M} \right)  2^{-\frac{B}{M-N}}
{M\!-\!1 \choose N\!-\!1}^{-\frac{1}{M-N}} \right)
\end{eqnarray*}
If we set this quantity equal to a desired rate gap $r > 0$
and solve for the required scaling of $B$ as a function of the
SNR (in dB) we get:
\begin{eqnarray} \label{eq-scaling_N}
B &=& \frac{M-N}{3} P_{dB} - (M-N) \log_2 c \\ &&
 - (M-N) \log_2 \left(\frac{M}{M \! - \! N \! + \! 1} \right) - \log_2  {M \! - \! 1 \choose N \! - \! 1} ,
\nonumber
\end{eqnarray}
where $c = 2^r \cdot e^{-(\sum_{l=M-N+1}^{M-1} \frac{1}{l})} - 1$.
Note that a per user rate gap of $r=1$ bps/Hz is equivalent to a 3 dB gap
in the sum rate curves.
If we compute the difference between this expression and 
the feedback load required when $N=1$ (given in (\ref{eq-scaling})) 
and a 3 dB gap is desired ($r=1$), we can get the
following approximation for the feedback reduction as a
function of the number of mobile antennas $N$:
\begin{eqnarray*}
\Delta_{FB}(N) \approx \frac{N-1}{3} P_{dB}  +\log_2  {M \! - \! 1 \choose N \! - \! 1}
- (N-1) \log_2 e.
\end{eqnarray*}
For $N=2$, the feedback savings is given by:
\begin{eqnarray*}
\Delta_{FB}(2) \approx \frac{1}{3} P_{dB}  + \log_2  (M-1)
- \log_2 e.
\end{eqnarray*}

\begin{figure}
    \centering
    \epsfig{file=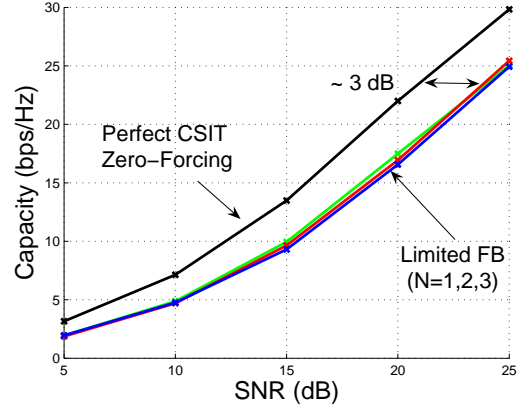, width=3.0in}
\caption{Sum rate of $M=K=6$ downlink channel}
   \label{fig-sum_rate}
\end{figure}

The sum rate of a 6 transmit antenna downlink channel is
plotted in Fig. \ref{fig-sum_rate}.  The perfect CSIT
zero-forcing curve is plotted along with the rates
achieved using finite rate feedback with the feedback
load scaled as specified in (\ref{eq-scaling_N}) for
$N=1,2$ and $3$.  Notice that the rates achieved for 
different numbers of transmit antennas are nearly
indistinguishable, and all three curves are approximately
3 dB shifts of the perfect CSIT curve.  In this system,
the feedback savings at 20 dB is 7 and 12 bits, respectively,
for $2$ and $3$ receive antennas.

\bibliographystyle{IEEEtran}
\bibliography{capacity}

\end{document}